# ACOUSTIC DETECTION IN SUPERCONDUCTING MAGNETS FOR PERFORMANCE CHARACTERIZATION AND DIAGNOSTICS

M. Marchevsky, X. Wang, G. Sabbi and S. Prestemon, LBNL, Berkeley CA, 94720, USA

*Abstract*

Quench diagnostics in superconducting accelerator magnets is essential for understanding performance limitations and improving magnet design. Applicability of the conventional quench diagnostics methods such as voltage taps or quench antennas is limited for long magnets or complex winding geometries, and alternative approaches are desirable. Here, we discuss acoustic sensing technique for detecting mechanical vibrations in superconducting magnets. Using LARP high-field $Nb_3Sn$ quadrupole HQ01 [1], we show how acoustic data is connected with voltage instabilities measured simultaneously in the magnet windings during provoked extractions and current ramps to quench. Instrumentation and data analysis techniques for acoustic sensing are reviewed.

## INTRODUCTION

Acoustic sensing of mechanical events in solids has a long history and has also been used in the past to access quench locations in superconducting magnets [2-9]. The advantages of this method are its non-intrusiveness, absence of sensitivity to magnetic field and use of inexpensive sensors that are easily adaptable to various magnet configurations. Sound propagation velocity of several km/s is typically faster than the quench propagation velocity; it allows for the mechanical detection to be accomplished on a millisecond time scale that is comparable (or faster) to other techniques. Furthermore, using acoustic sensor arrays, sound sources can be localized through triangulation with centimetre accuracy, and selectivity for different kinds of events can be achieved through post-processing and analysis in time and frequency domain.

Interpretation of the acoustic data is nevertheless a challenging problem. This is because sound in magnets can be generated by different mechanisms, most notably:
- sudden mechanical motion of a cable portion or coil part;
- cracking and/or fracture of epoxy;
- flux jump, as current re-distribution in the cable leads to the sudden local variation of the electromagnetic force;
- quench development, as formation of a hot spot leads to the quick thermal expansion and corresponding local stress build-up.

A common feature of all these sound-generating events is that they are usually associated with well-localized sources. Sound waves propagate radially from such source and eventually get reflected by the material boundaries, converted into resonant vibrational modes of the structure and into heat. Structural vibrations are of special importance for interpreting the sound signals, as various transverse (sound), longitudinal (bending) and more complex torsional modes can be excited by a single intrinsic mechanical event or just by the ambient background noise (helium boiling, cryostat vibrations, etc.); those resonances may then "ring" for a significant period of time (100-300 ms) due to relatively high (~100) mechanical quality factor of a typical magnet structure. The most interesting frequency range is the one associated with local vibration of a small component (cable, strand), and it is usually well above the range of structural mechanical resonances. Using high-pass filtering and post-processing one can therefore select the signal portion representative of a particular event and establish its precise origin and timing.

## INSTRUMENTATION

We have developed a system for acoustic sensing based on piezoelectric (PZT) transducers, cryogenic amplifiers and synchronous DAQ system. It was first tested using a room temperature arrangement and later used during the test of the LARP HQ01e magnet [1].

### Acoustic sensors and data acquisition

Piezoelectric transducers are widely used for acoustic sensing. In superconducting magnets, they are robust and sensitive to small structural vibrations: sensitivity to events as small as 0.2 µJ has been reported in [8]. In our system, we use disks of SM118 type piezoelectric ceramic, polarized across thickness with dimensions 10 mm outer diameter, 5 mm inner diameter, 2 mm thickness, and self-resonance frequency of (154 ± 4) kHz. (Fig. 1, left). Ring shape of the PZTs allowed for an easy installation on the magnet using a single set screw. In order to improve signal-to-noise ratio and eliminate need for using coaxial lines, we have combined our transducers with custom-built cryogenic amplifiers based on GaAs MOSFET and operating in the temperature range of 1.9-300 K; room-temperature gain of the amplifiers is ~ 3-5.

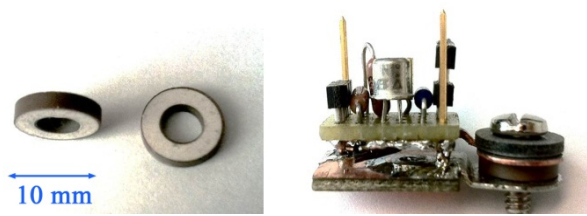

Figure 1: PZT transducers (left) and the cryogenic amplifier - PZT sensor assembly (right).

Each amplifier was interfaced to the room-temperature electronics using a twisted pair of wires and also battery-powered through that pair. The transducer-amplifier assembly is shown in Fig.1 (right). Fast DAQ (Yokogawa 7000) with simultaneous 1 MHz sampling was used to acquire data. LabView-based software was developed to perform signal frequency analysis and localize the sound source based on acoustic signal timing. Correct operation of the amplified piezo-sensors at 4.2 K was verified using a cryogenic insert to the transport helium dewar.

*Room temperature test*

Two sensor assemblies were installed on the HQ coil "endshoes", as shown in Fig. 2; distance between the sensors is $l_c$ = 0.96 m. Sounds were excited by slight knocking on the coil using a small metal key. By timing the difference between signal onsets, the sound velocity in the coil was measured as $v_s$ = (4.2 ± 0.1) km/s and the locations $x_k$ of the "knocks" were determined as

$$x_k = 0.5 \cdot (l_c + v_s \Delta t_{AB})$$

within ~ 50 mm accuracy.

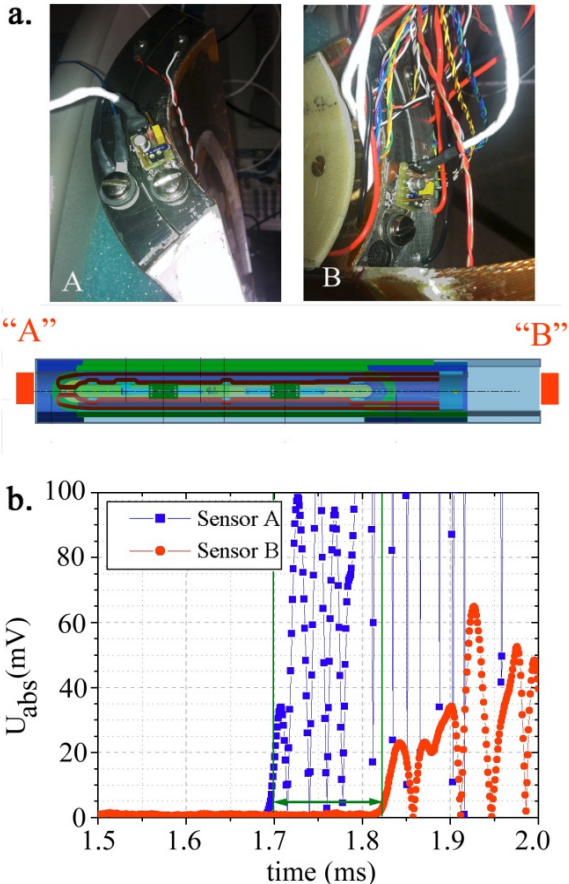

Figure 2: (a) Experimental arrangement for the room temperature test. Two PZT sensors ("A" and "B") were attached to the coil endshoes. (b) Typical signals measured upon slight "knocking" on the coil. Time difference $\Delta t_{AB}$ = 0.13 ms between the signal onsets (marked with an arrow) corresponds to the sound source location at 27 cm towards the sensor "A" from the center of the coil.

## EXPERIMENTAL RESULTS

*Installation on the magnet*

After the successful test at room temperature, two sensors were installed at the LARP HQ01 quadrupole $Nb_3Sn$ magnet [9] and tested at cryogenic temperature during magnet operation. The locations were chosen at the opposite sides of the magnet; one sensor was bolted to the magnet load plate and another one to the magnet shell; see Fig. 3.

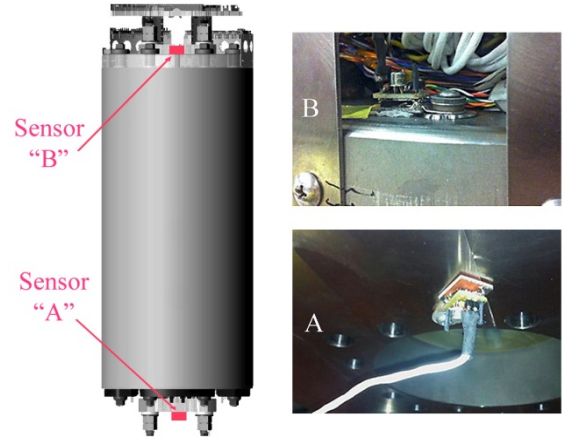

Figure 3: (left) HQ01e3 magnet on the stand with acoustic sensor locations marked with arrows. (right) amplified piezo-sensors bolted to the magnet shells (top) and the loadplate (bottom).

Typical mechanical resonance spectra of the magnet measured on the support stand using same technique as in the coil-on-the-table experiment reveals numerous peaks in the range of 0.3-10 kHz; associated with various compression, bending and torsional self-resonances of the magnet structure. Result for the power spectrum of the shell and load plate sensors are shown in Fig. 4.

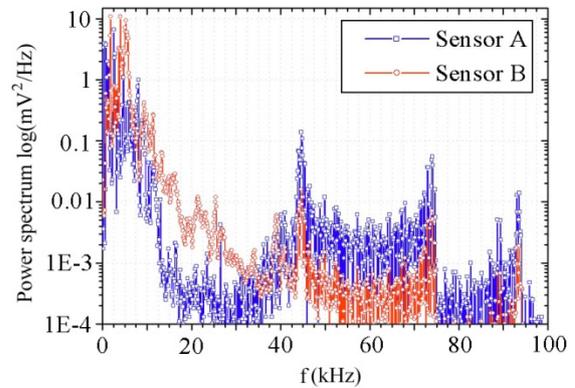

Figure 4: Power spectra of acoustic vibrations in the HQ01e3 magnet, measured in response to the mechanical excitation ("knocking") on the shell. Above ~10 kHz the resonant peaks are strongly suppressed in amplitude compared to those in the low-frequency range.

Note, that the spectra contain no significant peaks above 10 kHz. When the magnet shell was excited, strong signals were measured with both transducers. However, when the magnet interior (load-plates, rods) were excited, the response of the shell-mounted sensor "B" was much smaller compared to the sensor "A" response. This is indicative of the fact that mechanical vibrations in the coil structure remain fairly uncoupled of the shell excitations, which may be favorable for improved detections of the acoustic signals originating in the coils.

*Provoked extractions*

Upon cooling down, the magnet current was ramped up to 5.5 kA and a provoked extraction was triggered. Acoustic signals were recorded during the current ramp. First, we observed a significant noise associated with the current extraction. Acoustic waveforms obtained with the 5.5 kA provoked extraction are shown in Fig. 5. Clearly, the magnet is a good mechanical resonator with a quality factor Q~100, as the extraction-triggered "ringing" continues for nearly 1 s; this is ~4 times longer than the time constant of the magnet current transient. The spectral characteristics of the observed acoustic signals are similar to those seen with the room-temperature measurements, indicative of the "global" mechanical excitation of the entire magnet structure with the changing Lorentz force; no significant high-frequency sounds potentially indicative of the vibrating small parts were detected.

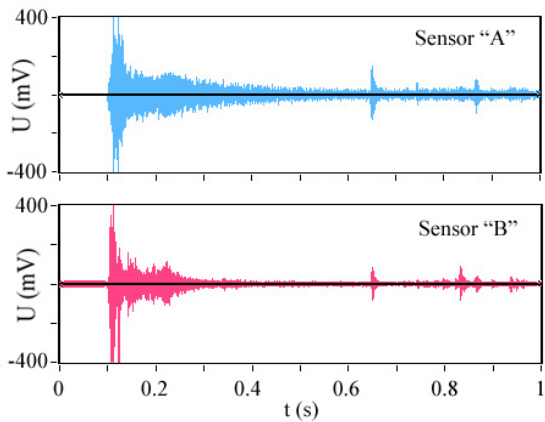

Figure 5: Sound emission waveforms resulting from the provoked extraction of the magnet current at 5.5 kA.

*Ramps to quench*

In the following, the magnet current was
- ramped up at 75 A/s to 9 kA, current was held steady for 3 min and then ramped back down to zero;
- ramped up at 75 A/s to a spontaneous quench, that occurred at 10.87 kA.

In all these experiments, voltage imbalance (formed by subtracting voltage of two halves of the magnet, usually employed to detect quenches) was recorded simultaneously with the acoustic signals. Data recording rate was 1 MHz and the time window width is 0.2 s. Acquisitions were triggered whenever *either* imbalance or sound was detected to exceed a threshold level; for acoustic signal the threshold was chosen at 5 mV and for the imbalance at 75 mV.

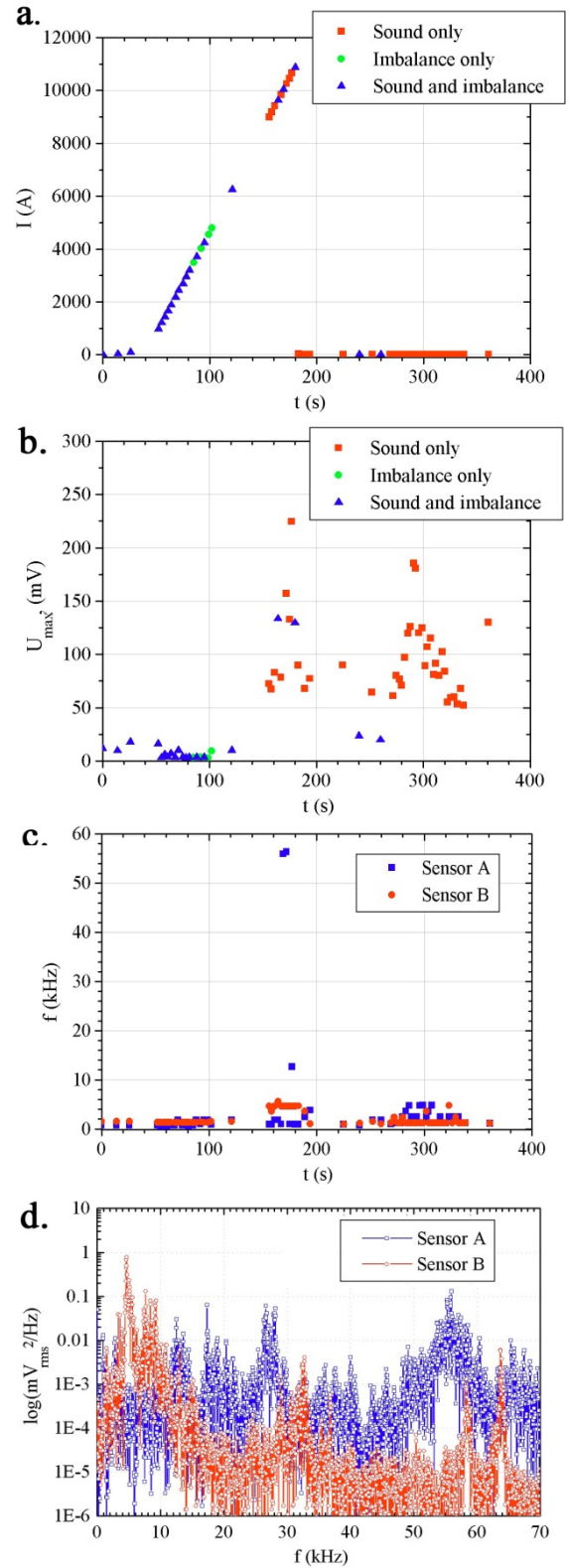

Figure 6: Summary of events triggered by either sound emissions or imbalance variations during the 75 A/s ramp to quench at 10870 A; each point represents a single acquisition cycle of 0.2 s. Time dependence of current (a), maximal sound amplitude (b) and maximal detected sound frequency (c) are shown. (d) Acoustic spectra corresponding to the highest magnet current (~10800 A).

Results of the spontaneous quench ramp are shown in Fig. 6. Four possible types of events were identified:
  Below 5 kA:
- Imbalance variation without any associated sound;
- Imbalance variation associated with weak sound signals.

  At 8.5 kA and above:
- Stronger sounds with no association with imbalance variations.

  Around 10-10.5 kA:
- Stronger sounds associated with imbalance "spikes".

The low-current imbalance variations are known to be caused by flux jumps in the superconducting cable and have been observed in the earlier tests of HQ [1].

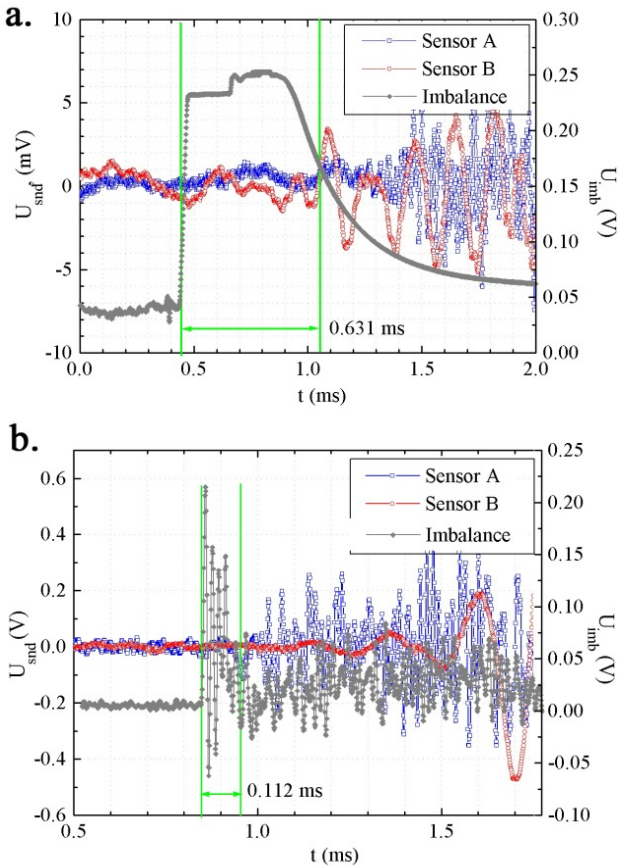

Figure 7: Simultaneously acquired imbalance and acoustic signals at magnet currents of 2440 A (a) and 10036 A (b). In (a) the 0.63 ms delay between the imbalance onset and the sound would place the sound source ~2.6 m away from the "A" transducer, which would be outside of the magnet. In (b) the 0.11 ms delay corresponds to ~0.46 m distance, hence sound is produced within the magnet length.

Our measurements show that the weak acoustic emissions are associated with at least some of these events; this result is consistent with the earlier studies [4, 5].

What is most interesting, however, is that the much stronger sounds are recorded at higher currents where flux jumps are absent. Moreover, these stronger sounds are also associated with much higher frequencies (40-60 kHz) than those observed in provoked extraction experiments. In Fig. 6 (d) the power spectrum of the acoustic signal of sensor "A" (attached to the loadplate) shows an absolute peak at ~56 kHz that becomes prominent only at magnet currents above ~ 9 kA. Same high-frequency sound was detected in other 75 A/s current ramps to 9 kA and back, but without quenching.

To understand origin of the observed acoustic emissions, we have attempted to determine sound source locations for various triggered event, at low and high currents. In Fig. 7 two results of noise source localization are shown for the magnet current of 2440 A (plot a) and 10036 A (plot b) respectively. It turns out, that, based on the signal timing, the sound source in (a) would be located outside the magnet. Such result suggests that in reality, there could be a delay between the flux jump onset (seen as imbalance variation) and the sound generation. On the other hand, in (b) the sound is produced within the magnet length and also the imbalance exhibits multiple fast fluctuations simultaneously with the sound. This observation seems most consistent with the mechanical event [8], such as stick-slip motion of the superconducting cable or supporting structure that is responsible for both sound and the imbalance "spike".

## CONCLUSIONS

Amplified piezo-sensors, in combination with fast data acquisition and processing techniques show good potential for real-time characterization of various mechanical events in superconducting magnets during ramping, quench and recovery. We have shown that acoustic signals generated by flux jumps and mechanical motion events in the superconducting accelerator magnet have distinctly different features. HQ magnet exhibits occasional weak acoustic emissions correlated with the flux jumps below 5 kA as well as the larger amplitude high-frequency (>50 kHz) emissions unrelated to flux jumps and only seen above 9 kA. The sounds recorded at high current are occasionally correlated with the short spikes in the magnet electrical imbalance and multiple fast fluctuations most likely caused by stick-slip motion of the conductor.

Further development of the acoustic technique is needed, focusing on improving sensitivity and selectivity to small signals, developing instrumentation and software for precise localization of the sound sources and quantifying energy release in the detected acoustical events. We also plan to access feasibility of the full-scale acoustic quench detection and diagnostic system in the upcoming magnet tests.

## ACKNOWLEDGMENT

Technical assistance from P. K. Roy, M. Turqueti, T. Lipton and R. Albright is gratefully acknowledged.

This work is supported by supported by the Director, Office of Science, High Energy Physics, U.S. Department of Energy under contract No. DE-AC02-05CH11231.

## REFERENCES

[1] M. Marchevsky, G. Ambrosio, B. Bingham, R. Bossert, S. Caspi, D. W. Cheng, G. Chlachidze, D. Dietderich, J. DiMarco, H. Felice, P. Ferracin, A. Ghosh, A. R. Hafalia, J. Joseph, J. Lizarazo, G. Sabbi, J. Schmalzle, P. Wanderer, X. Wang, A. V. Zlobin, Test of HQ01, a 120 mm Bore LARP Quadrupole for the LHC Upgrade, IEEE Trans. Appl. Supercond. 22, 4702005 (2012), and ref. therein.

[2] P.P. Gillis, Dislocation motion and acoustic emission, ASTM STP 505, 20-29, 1972.

[3] G. Pasztor and C. Schmidt, Dynamic stress effects in technical superconductors and the "training" problem of superconducting magnets, J. Appl. Phys. 49, 886 (1978).

[4] H. Brechna and P. Turowski, Training and degradation phenomena in superconducting magnets, Proc. 6th Intl. Conf. Magnet Tech. (MT6) (ALFA, Bratislava, Czechoslovakia) 597, (1978).

[5] G. Pasztor and C. Schmidt, Acoustic emission from NbTi superconductors during flux jump, Cryogenics 19, 608 (1979).

[6] O. Tsukamoto and Y. Iwasa, Sources of acoustic emission in superconducting magnets, J. Appl. Phys. 54, 997 (1983).

[7] M. Pappe, Discussion on acoustic emission of a superconducting solenoid, IEEE Trans. on Magn., 19, 1086 (1983).

[8] O.O. Ige, A.D. McInturf and Y. Iwasa, Acoustic emission monitoring results from a Fermi dipole, Cryogenics 26, 131, (1986).

[9] Y. Iwasa, Mechanical disturbances in superconducting magnets-a review, IEEE Trans. on Magn., 28, 113 (1992).